\documentclass[12pt]{iopart}
\usepackage{iopams}
\usepackage{epsfig}
\usepackage{citesort}
\usepackage[dvips]{color}

\def\aj{AJ}

\def\apj{ApJ}
\def\apjl{ApJ}

\def\aap{A\&A}

\def\mnras{MNRAS}

\def\prd{Phys.~Rev.~D}

\newcommand{\hMpc}{h^{-1}\, \rm Mpc}
\newcommand{\kms}{\textrm{\,km s}^{-1}}
\newcommand{\vb}{\mathbf{v}}
\newcommand{\Om}{\Omega_m}
\newcommand{\kb}{\mathbf{k}}

\definecolor{Black}{named}{Black}
\definecolor{Red}{named}{Red}

\newcommand{\fig}[1]{figure \ \ref{#1}}
\newcommand{\Fig}[1]{Figure \ref{#1}}
\newcommand{\Eq}[1]{Eq.\ \ref{#1}}
\begin{document}

\title{Precision measurements of large scale structure with future type Ia supernova surveys}

\author{Steen Hannestad, Troels Haugb{\o}lle, Bjarne Thomsen}
\address{Department of Physics and Astronomy, University of Aarhus, Ny Munkegade, DK-8000 Aarhus C, Denmark}

\ead{\mailto{sth@phys.au.dk}}
\date{{\today}}

\begin{abstract}
Type Ia supernovae are currently the best known standard candles at
cosmological distances. In addition to providing a powerful probe of
dark energy they are an ideal source of information about the
peculiar velocity field of the local universe. Even with the very
small number of supernovae presently available it
has been possible to measure the dipole and quadrupole of the local
velocity field out to $z \sim 0.025$. With future continuous all-sky
surveys like the LSST project the luminosity distances of tens of
thousands of nearby supernovae will be measured accurately. This
will allow for a determination of the local velocity structure of
the universe as a function of redshift with unprecedented accuracy,
provided the redshifts of the host galaxies are known. Using
catalogues of mock surveys we estimate that future low redshift
supernova surveys will be able to probe $\sigma_8$ to a precision of
roughly 5\% at 95\% C.L. This is comparable to the precision in
future galaxy and weak lensing surveys and with a relatively modest
observational effort it will provide a crucial cross-check on future
measurements of the matter power spectrum.
\end{abstract}
%\pacs{14.60.Pq,95.35.+d,98.80.-k}
\maketitle

%%%%%%%%%%%%%%%%%%%%%%%%%%%%%%%%%%%%%%%%%%%%%%%%%%%%%%%%%%%%%%%%%%%%%%
\section{Introduction} %%%%%%%%%%%%%%%%%%%%%%%%%%%%%%%%%%%%%%%%%%%%%%%
%%%%%%%%%%%%%%%%%%%%%%%%%%%%%%%%%%%%%%%%%%%%%%%%%%%%%%%%%%%%%%%%%%%%%%
Precision measurements of distant type Ia supernovae have been an
essential ingredient in building the present standard model of
cosmology in which the energy density of the universe is dominated
by dark energy \cite{perlmutter99,schmidt98,riess98}. The main
interest so far has been in supernovae in the redshift range 0.5-1.5
\cite{tonry03,knop03,riess04,astier06,Wood-Vasey:2007jb,riess07}, which corresponds
to the transition region between matter and dark energy domination.

However, there is a rapidly increasing interest in supernovae at
much lower redshifts
\cite{Bonvin:2006en,Haugboelle:2006uc,Hui:2005nm,Jha:2007,neill,Wang:2007nh}.
The low redshift supernovae provide an anchor
for the Hubble diagram which is important for the determination of
dark energy parameters. Since they are much brighter they are also
useful as a laboratory for calibrating the relation between the
supernova light curve shape and its inherent brightness.

A potentially even more interesting feature of low redshift
supernovae is that the velocity field associated with the large
scale structure of the universe is imprinted on the apparent
magnitudes and redshifts of the supernovae. If this is not corrected for it will
lead to relatively large and correlated errors in the measured
luminosity distance of low redshift supernovae, a potentially
serious source of error for future dark energy surveys
\cite{sugiura:1999,Bonvin:2005ps,Bonvin:2006en,Hui:2005nm,neill}.
However, the velocity field can also be used to probe the local
large scale structure of the universe.

With the current fairly small datasets at low redshift only the
local dipole \cite{tonry:2000,hudson2} and quadrupole
\cite{Haugboelle:2006uc} terms in the velocity field have been
measured (see also \cite{Wang:2007nh}), and agree with other
types of flow measurements. These preliminary studies, however,
clearly demonstrate the potential of using type Ia supernovae for this purpose. The very small
intrinsic scatter in luminosity means that the local velocity field
can be measured as precisely with $\sim$100 supernovae as with an
order of magnitude more galaxies used as standard candles.

In the present work we will, for the first time, study how future
type Ia supernova measurements can be used for detailed studies of
the large scale structure of the universe. It turns out that the
large scale velocity power spectrum can be measured quite precisely
using around $10^5$ supernovae, a number which will be observed with
surveys such as the Large Synoptic Survey Telescope (LSST)
\cite{lsst}.

In order to investigate the potential of such future measurements we
use detailed mock supernova catalogues produced from high resolution
N-body simulations. We find that the survey strategy for measuring
the supernovae is very important and investigate this phenomenon in
detail.

Our most important finding is that future supernova surveys can be
used to probe cosmological parameters such as $\sigma_8$, the
amplitude of fluctuations on small scales, at a precision which is
competitive with other future cosmological probes.

In Section 2 we present the assumed supernova rates and discuss
upcoming surveys in this context. Section 3 provides a review of the
velocity power spectrum, as well as a simple method for analytically
calculating the angular velocity power spectrum as a function of
redshift. Details of the simulations and the power spectrum
extraction are provided in Section 4. The extraction of cosmological
parameters from our simulated data sets is discussed in Section 5,
and finally Section 6 contains our conclusions.

%%%%%%%%%%%%%%%%%%%%%%%%%%%%%%%%%%%%%%%%%%%%%%%%%%%%%%%%%%%%%%%%%%%%%%
\section{The supernova rate and future surveys} %%%%%%%%%%%%%%%%%%%%%%
%%%%%%%%%%%%%%%%%%%%%%%%%%%%%%%%%%%%%%%%%%%%%%%%%%%%%%%%%%%%%%%%%%%%%%
The local type Ia supernova rate is roughly
\cite{Sullivan:2006,rate1,rate2}
\begin{equation}
\label{eq:rate}
R \simeq 1.2^{+2}_{-1} \times 10^{-4} \, {\rm
yr}^{-1} \, h^3 \, {\rm Mpc}^{-3}\,,
\end{equation}
and while there is considerable uncertainty about the exact rate
(see e.g.\ \cite{rate} for a 3 times higher rate deduced using
SNfactory data), we will adopt the above rate. If the actual rate is
lower (higher) then simply a longer (shorter) survey time would be
needed to reach the same number of measured supernovae.

Out to a redshift of $z_{\rm max}$, where $z_{\rm max} \ll 1$ the
total number of supernovae per time is then
\begin{equation}
N = 1.3 \times 10^4 \left(\frac{z_{\rm max}}{0.1}\right)^3 \, {\rm yr}^{-1}
\end{equation}
To probe the velocity structure on different length scales, we will
analyse the Supernovae in redshift bins. We compare with theory by calculating
the angular power spectrum, $C_l$, in each redshift bin.

As will be seen below, using upcoming surveys and
collecting data in a three-year period, taking a conservative
approach, the maximum multipole which can be probed at $z<0.13$ is
$l_{max} \sim 20$. However, the number of supernovae in a bin increases
cubically, while the error on individual supernovae $\delta v_r$ increases
linearly with distance. The overall error goes like $\delta
v_r/N_{bin}^{1/2}$ or $z^{-1/2}$, and therefore the precision, that
the radial velocity of a given {\em angular} scale can be measured
by, increases with distance.

We also note that exactly because the effective error scales like
$\delta v_r/N_{bin}^{1/2}$, an improved understanding of the light
curve--absolute magnitude relation, and environmental effects, such as
dust extinction could dramatically decrease the
error on the angular power spectrum, compared to what is forecasted
in this paper. Such a decrease is not unreasonable to expect, given
the much larger and more homogeneous sample of low redshift type Ia
supernovae that will be available in the near future.

\subsection{Future surveys}
Several upcoming surveys will map large portions of the sky at
regular intervals to look for either transient sources or hitherto
undiscovered solar system objects. In the near future there will be
projects such as Pan-STARRS \cite{panstarrs} and SkyMapper
\cite{skymapper}, and in about 2013 the LSST project \cite{lsst}
will start.

The scan strategy of these surveys is not necessarily optimised to
search for nearby supernovae, while all of them have dedicated
programs for intermediate or high redshift supernovae. Despite this,
with the LSST a very large fraction of all type Ia supernovae out to
redshifts of order 0.2 will be detected and their light curves
measured in detail.

However, since none of these experiments will perform follow-up
spectroscopy of the host galaxy, this task will have to be
undertaken using other telescopes. Fortunately the host galaxies
will usually be sufficiently bright for spectroscopy even with
modest sized telescopes, and the task is therefore not unmanageable.
We note that a sufficiently accurate spectrum can be obtained in
roughly 20 minutes using a 1m class telescope \footnote{Based on an
assumed accuracy of 100 $\times (z / 0.05)\, \kms$ and a redshift of
$\sim 0.1$.}, and thousands of redshifts per year can therefore be measured
relatively easily using just a few such telescopes. We also note
that the important part of the spectroscopy is to determine the redshift
of the host galaxy, not the supernova, and that the spectroscopic
observations can therefore in principle be done long after the
supernova event itself.

Of course this is true only if the supernova can be determined to be
a type Ia from the light curve measurements alone with very close to
100\% efficiency. With a precise light curve for a supernova in 6 filters
sampled every fourth night in a rolling search, it should be possible to reliably
detect if it is a type Ia supernova using only photometry
\cite{Wang:2007,Kuznetsova:2006}, and determine the
luminosity distance very precisely.

In the following we will assume full sky coverage and use numbers
corresponding to the total number of supernovae occurring in three
years (assuming the rate given in \Eq{eq:rate}). If the
number of supernovae is lower the numbers can easily be scaled to a
longer observation period. Full sky coverage is not essential as
long as it is significantly larger than 0.5, and the unobserved area
is distributed in at most a few large patches, as is also true for
CMB observations.

%%%%%%%%%%%%%%%%%%%%%%%%%%%%%%%%%%%%%%%%%%%%%%%%%%%%%%%%%%%%%%%%%%%%%%
\section{The angular velocity power spectrum} %%%%%%%%%%%%%%%%%%%%%%%%%%%%%%%%%%
%%%%%%%%%%%%%%%%%%%%%%%%%%%%%%%%%%%%%%%%%%%%%%%%%%%%%%%%%%%%%%%%%%%%%%
In order to make the analysis of large scale velocity flows simpler
we will work with the angular velocity power spectrum. This in turn
also makes the extraction of cosmological parameters from the
synthetic data easier. Following \cite{Peebles:1993,Verde:2006} the
angular velocity power spectrum can be simply related to the matter
power spectrum in the linear limit. The continuity equation of mass
relate the velocity and density: $\nabla \cdot \vb = -a \partial
\delta /
\partial t$. Fourier transforming it, we get
\begin{equation}
\kb \cdot \vb_k = - \rmi H f(\Omega) \delta_k \,,
\end{equation}
where $\delta_\kb$ is the Fourier transform of the density contrast,
and $f(\Omega) =\rmd \log \delta/\rmd\log a \sim \Om^{0.6}$ is the
velocity growth factor \cite{peebles1976}.

The peculiar velocities are measured as averaged quantities over a
certain scale. For example the peculiar velocity derived from observing
a supernova, and the host galaxy redshift, is the smoothed velocity
on the scale of the host galaxy. Likewise, in a N-body simulation the
velocity field is measured using a smoothing kernel. Therefore the
power spectrum, be it synthetic or observed, is smoothed by a window
function, and related to the matter power spectrum as
\begin{equation}
P_v(k) = H^2 f(\Omega)^2 k^{-2} P_m(k) |W(kR)|^2\,
\end{equation}
where $W(kR)$ is the Fourier transform of the window function.

What we are measuring is the angular power spectrum of the radial
peculiar velocities on a shell. It can be related to the 3D velocity
power spectrum as \cite{Verde:2006}
\begin{equation}\label{eq:clv}
C_l = \frac{4\pi}{2l+1}(l\mathcal{B}_{l-1}+(l+1)\mathcal{B}_{l+1})\, ,
\end{equation}
where
\begin{equation}
\label{eq:bl} \mathcal{B}_l = 4\pi\int\frac{k^2 \rmd
k}{(2\pi)^3}P_v(k) j_l(k x)^2\, ,
\end{equation}
$j_l(k x)$ is the spherical Bessel function, and $x$ is the comoving
distance to the shell. \Fig{fig:fig2} shows the analytic angular
power spectra at different distances compared to the power spectra
obtained by directly finding the radial velocity field averaged over
different observers in a large scale N-body simulation. There is an
excellent agreement between the synthetic spectra and the model for
small $l$, while at large $l$ the discrepancy is due to the
assumption of a linear power spectrum, and no shell crossing in
Fourier space.

To be consistent with the N-body simulation the radial velocity field in
the simulation is determined by smoothing the radial velocities
of the particles with the same adaptive smoothing kernel of Monaghan and
Lattanzio \cite{Monaghan:1985} with 33 neighbours that is used in the simulation.
The form of the window function for this smoothing kernel is not a priori clear, since the
kernel size is adaptive and varies with density. Therefore one cannot just Fourier
transform the kernel with a fixed $R$. Nonetheless we have found empirically that
\begin{eqnarray}
|W(kR)|^2 = \frac{1}{1 + kR}\,, \\
R = 2 L_{box} \left(\frac{N_{neighbour}}{N_{particles}}\right)^{1/3}
\end{eqnarray}
is an excellent approximation. Here $L_{box}$ is the box size, $N_{particles}$
the number of particles in the simulation, and $N_{neighbour}$ the number of
neighbour particles used in the smoothing process.

The analytic matter power spectrum $P_m(k)$ is a standard linear matter power spectrum with
$(h,\Omega_m,\Omega_\Lambda,\sigma_8)=(0.7,0.3,0.7,0.9)$ computed
using CAMB \cite{CAMB}.

\subsection{Analytic Approximation}
From \Eq{eq:clv} we see that the velocity angular power spectrum is
related to the matter power spectrum through a convolution. From
Eqs.~(\ref{eq:clv}) and (\ref{eq:bl}) we can write the angular
power spectrum as
\begin{equation}
C_l = \int k^2 dk P_v(k) \Delta^2(k),
\end{equation}
with
\begin{equation}
\Delta^2(k)=\frac{2}{\pi(2l+1)}\left(lj_{l-1}^2(kx) +
(l+1)j_{l+1}^2(kx)\right).
\end{equation}

\begin{figure}
\begin{center}
\includegraphics[width=0.8 \textwidth]{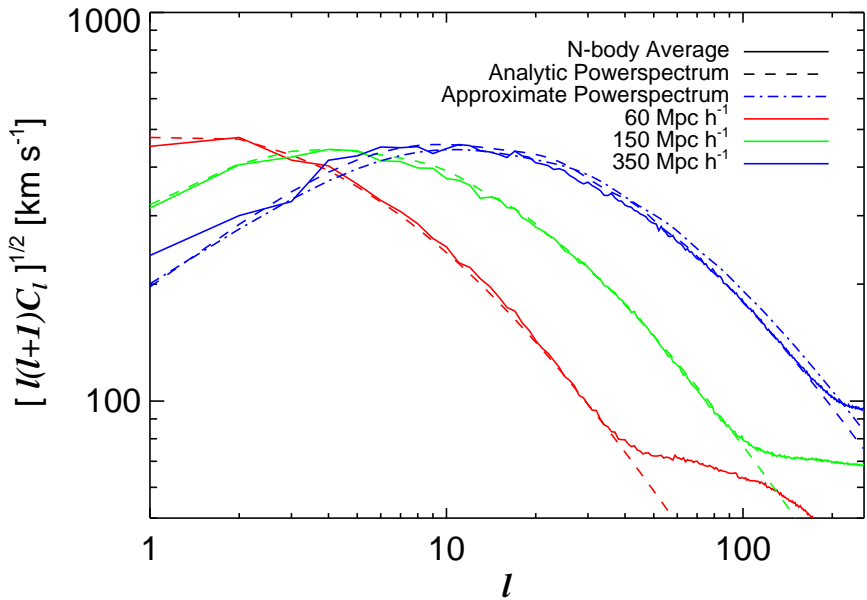}
\caption{The angular velocity power spectrum at different comoving distances
based on an ensemble average from the N-body simulations (full line), and
the analytic model (dashed line). The approximation (\Eq{eq:clvsimple}) to
the full analytic model is also shown as the dot-dashed line for $350\, \hMpc$.}\label{fig:fig2}
\end{center}
\end{figure}

Following the approach of \cite{Oystein:2003}, the two $k$-dependent
terms in the convolution can be significantly simplified in the
limit $l\gg1$, where the integral is bound from below.
The Bessel function is rapidly oscillating, but the envelope has a simple analytic form
\begin{eqnarray}
\Delta^2_l(k) = \left\{ \begin{array}{ll}
\frac{\rme^2}{8(2l+1)l}\left(\frac{ekx}{2l}\right)^{2(l-1)} \sim 0 & ,\, k < k_* \\
\frac{1}{\pi(k x)^2} &,\, k \ge k_*
\end{array}\right. \\
|W(kR)|^2 = \left\{ \begin{array}{ll}
1 & , \, kR < 1 \\
(kR)^{-1} & , \, kR \ge 1
\end{array}\right\} \simeq \frac{1}{1+(kR)}\,,
\end{eqnarray}
where $k_*=\frac{2}{\rme} [(2l+1)/l]^{1/2l} l/x\simeq \frac{2}{\rme}
(1 - \frac{\ln 2}{2 l}) l/x$. Collecting terms we find simply
\begin{equation}\label{eq:clvsimple}
C_l = \frac{H^2 f(\Omega)^2}{\pi}  \left[ \int^{1/R}_{k_*} \frac{dk}{(k x)^2} P_m(k)
 + \int^{\infty}_{1/R} \frac{dk}{(k x)^2 (kR)} P_m(k) \right]
\end{equation}
A comparison between the exact formula \Eq{eq:clv} (the dashed line) and the
approximation \Eq{eq:clvsimple} (the dot-dashed line) is shown in \fig{fig:fig2},
and very good agreement is found. The Bessel
function acts like a triangular filter, such that power at a given
$l$ only is dependent on scales smaller than $\frac{2}{\rme}l/x$,
and the slow decline of $j_l(k x)^2$ for $\frac{2}{\rme}k>l/x$
washes out any sharp features and breaks in $P_m$
\cite{Oystein:2003}.

In order to get a simple analytic understanding of the behaviour of
$C_l$ we use a crude approximation for $P_m(k)$,
\begin{equation}
\label{eq:pspec}
 P_m(k) \sim \frac{A k}{1+B k^\alpha},
\end{equation}
where $\alpha=3$ provides a reasonable fit near the maximum of $P_m$
(although not the correct asymptotic limit for $k \to \infty$). Using this, one finds that for
$l \gg 1$
\begin{equation}
C_l \sim \log\left(\frac{x^3 + B l^3}{B l^3}\right)=\log
(1+l_0^3/l^3),
\end{equation}
where $l_0 = x/B^{1/3}$. From this one finds that $C_l$ has a
maximum at $l/l_0 = 0.89$ and therefore that the maximum for $C_l$
is proportional to $x/B^{1/3}$. $B$ can be related to the maximum
of $P_m(k)$ which is at $k=k_{\rm max}=1/(2B^{1/3})$. Therefore the
maximum in $C_l$ is directly proportional to the maximum in $P_m(k)$
and the comoving distance $x$ as long as $l_0 \gg 1$.

\subsection{Connecting the angular velocity power spectrum with the matter power spectrum}
A full-sky supernova sample has both drawbacks and advantages
compared with possible galaxy redshift surveys for probing the large
scale structure of the universe. The major advantage is that there
is no need to know the completeness function of the survey very
accurately and that the effective survey volume can be very large
using only a relatively small number of objects. However, this also
means that it is difficult to probe features on small scales because
of the sparseness of the sample. In this regard supernova surveys
are fairly similar to surveys like the Sloan Digital Sky Survey
Luminous Red Galaxy survey (SDSS-LRG) \cite{sdsslrg} in which a
relatively sparse sample of very bright galaxies is used.

The velocity survey is not affected by bias between baryonic and
dark matter, because it is the full matter distribution that sources
the velocity field. Together with the lack of sensitivity to
completeness it means that supernova velocity surveys are excellent
probes of the amplitude of fluctuations, i.e.\ $\sigma_8$. Given
enough statistics they can also be used to probe the shape
parameter, $\Gamma$, of the matter power spectrum which is related
to the $\alpha$ and $B$ parameters in \Eq{eq:pspec}, and to $\Omega_m$
(see \cite{Watkins:2007qn} for a first attempt at probing $\Gamma$ using
present supernova data).

In section \ref{sec:cosmo} we study the sensitivity to $\sigma_8$ and $\Omega_m$
of mock supernova surveys and find in particular that $\sigma_8$ can be
very well constrained.

%%%%%%%%%%%%%%%%%%%%%%%%%%%%%%%%%%%%%%%%%%%%
\section{Multipole analysis, survey strategy, and construction of mock catalogues}
%%%%%%%%%%%%%%%%%%%%%%%%%%%%%%%%%%%%%%%%%%%%
To make a forecast of what can be measured with future supernova catalogues we construct
mock catalogues using a refined version of the formalism in
\cite{Haugboelle:2006uc}. First, we generate a sample with $N_{tot}$ supernovae, and distribute
them in a redshift bins. The supernovae are sampled proportionally to density in a
large scale N-body simulation (in \cite{Haugboelle:2006uc} it was shown that
the results are only weakly dependent on the exact distribution as
long as it is semi-random). From this simulated dataset we calculate
the angular velocity power spectrum in the different redshift bins from
weighted least squares fits of spherical harmonic functions to the
radial velocities using the method of singular value decomposition.
The multipole expansion is terminated at $l_{max} = \max(\sqrt{N/3},20)$,
in order to keep the number of degrees of freedom larger than the number
of fitted coefficients, $(l_{max}+1)^2$, (and to keep the solution
practical in terms of CPU time).

To get a handle on cosmic variance we repeat the procedure
40 times for differently located observers, and to understand how
noise affect the measurement we add a Gaussian noise of $\Delta
m=0.08$ to the magnitude of the supernovae and redo the analysis 900 times
per observer.

In \fig{fig:fig2} we show the volume averaged angular power spectrum
calculated from the N-body simulation on a thin shell using Healpix
\cite{Gorski:2005}.
This corresponds to what one would find for an infinitely large supernova
sample, assuming that there are no errors on the individual supernovae,
and that they all are placed at exactly the same distance.

\subsection{The error budget}
\begin{figure}
\begin{center}
\includegraphics[width=0.8 \textwidth]{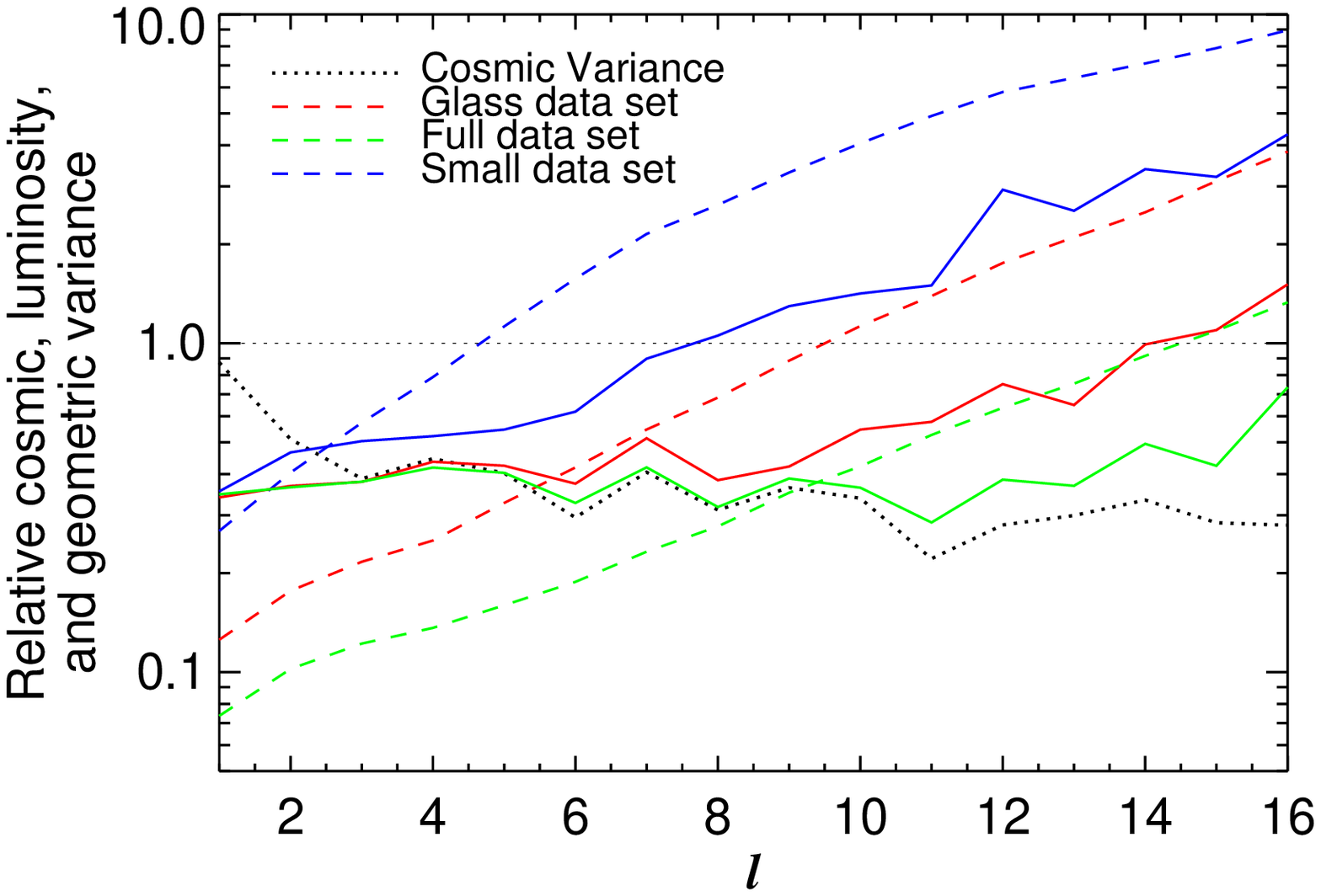}\\
\includegraphics[width=0.8 \textwidth]{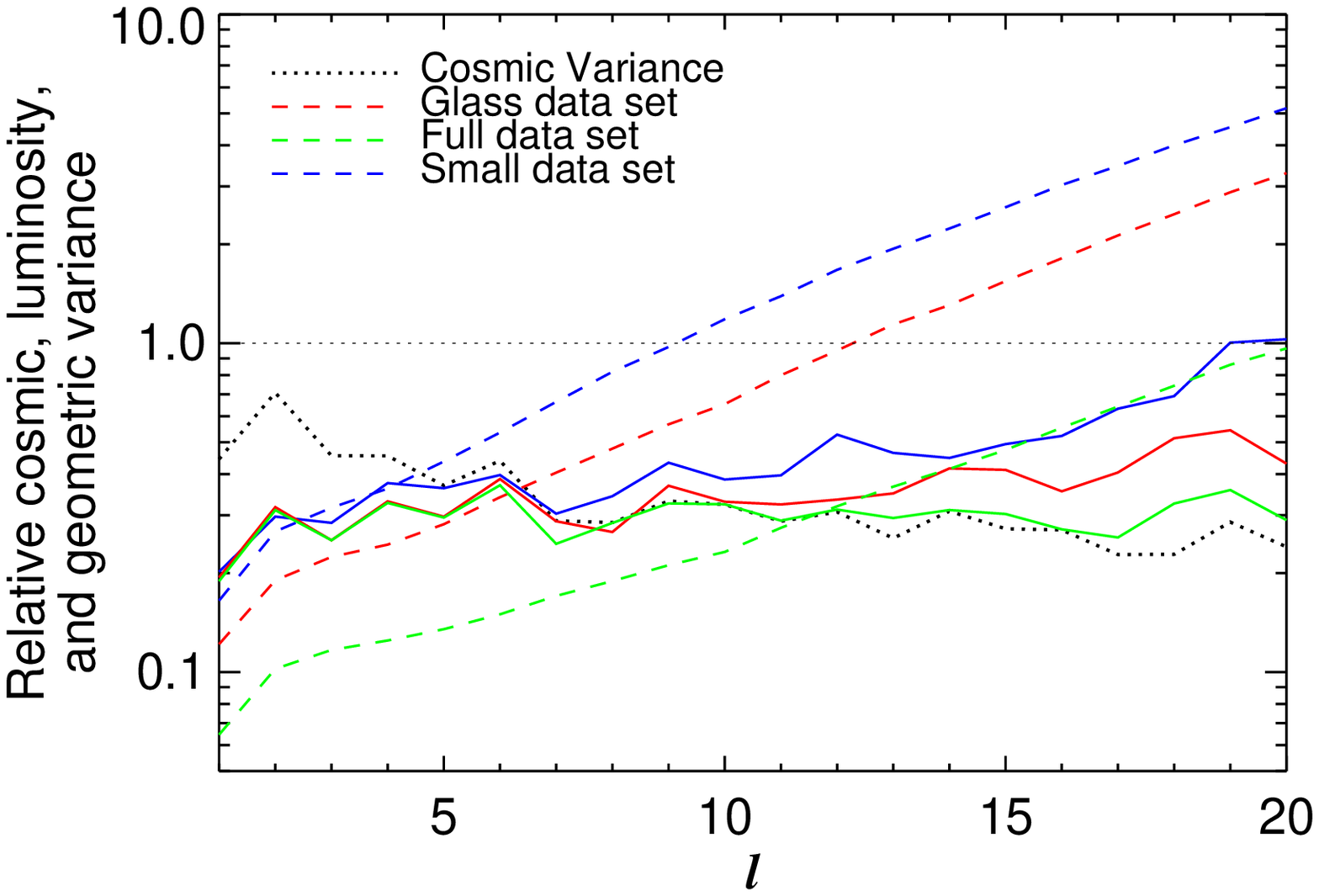}
\caption{The variances divided by the signal
in a redshift bin for the different error
terms. The upper (lower) panel is for a bin centered on
$100\,\hMpc$ ($150\,\hMpc$).
Dashed and full lines are for luminosity
and geometric errors respectively, while the thick dotted line is the
cosmic variance. The $100\,\hMpc$ bin
contains 752 supernovae in the small and glass data sets and 3009
supernovae in the full data set. The corresponding numbers for the
$150\,\hMpc$ bin are 1781 and 7125.}\label{fig:fig3}
\end{center}
\end{figure}

When comparing the true ensemble averaged angular power spectrum
with that derived from a specific finite sample of supernovae,
various errors are introduced. There is a \emph{luminosity error}
from the combined error in both apparent and absolute magnitudes of
the individual supernovae. Then there is a difference between an
angular power spectrum, sampled by supernovae in a redshift bin of
finite thickness, even in the absence of errors, and the true
underlying angular power spectrum. We call it the \emph{geometric error}.
It depends only on the specific geometry, or angular distribution, of
the supernova sample, and the thickness of the bin. Finally, at low
$l$ and small distances the error will be dominated by \emph{cosmic
variance} (see \fig{fig:fig3} for a breakdown of the error budget in two specific
redshift bins).

To understand the statistical properties, and compensate for the
noise, we make two independent set of mock catalogues of
observations with different synthetic observers. The first is used to draw
mock observers, while the second is used to model the noise, and
subtract it from the mock observers. This is done to mimic real
observations, where there are no correlations between the
observations and the mock catalogues used to model the noise. For
each observer in the second set of catalogues we measure the true
angular power spectrum, and the angular power spectra derived from
a given number of supernovae, with
and without noise included. This gives us a robust prediction for
the angular power spectra of the luminosity error and the
geometric error, as well as the variances on the two types of
errors. Comparing volume averaged underlying power spectra from
different observers we find the scatter from cosmic variance.
In figure \ref{fig:fig3} we show the three different types of error
terms for the three different mock surveys, which will be described in
detail below. It can be seen that the luminosity error eventually
begins to dominate at high $l$, while the geometric error is mostly
important when considering small number of supernovae. At low $l$
the error from cosmic variance dominates completely, as expected.

\begin{figure}
\begin{center}
\includegraphics[width=0.8 \textwidth]{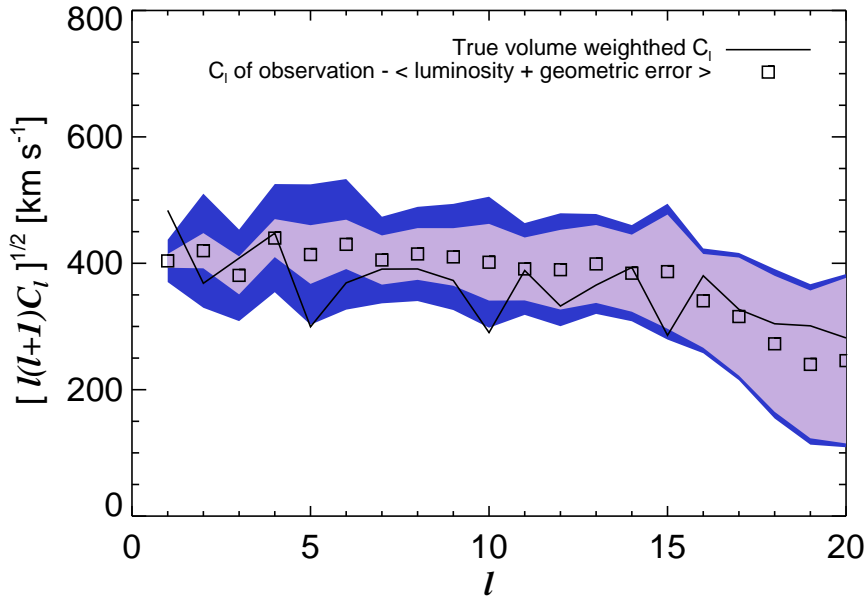}
\caption{
The radial velocity angular amplitude spectrum in a bin centered at a comoving distance
of $150 \,\hMpc$ based on a synthetic supernova survey bin with 7125 randomly distributed
supernovae.
Blue (purple) contours indicate the errors due to the noise correction
procedure and including (excluding) cosmic variance with a Gaussian
intrinsic and observational scatter of $\Delta m = 0.08$ on each supernova.
The full line is the true volume averaged amplitude spectrum for the specific observer,
while the squares show the corresponding
synthetic observation.}\label{fig:RNG7125_149}
\end{center}
\end{figure}

\subsection{Geometric errors and supernovae on a glass}
When considering a mock observation, for very low $l$ the true angular power spectrum
can easily be recovered. However, for large $l$ the geometric error is considerable. This
phenomenon occurs because there are ``holes'' in the sky coverage (this is discussed in
detail in \cite{Haugboelle:2006uc}). These holes lead to leaks of power between
different $l$ modes (the spherical harmonics restricted to the positions of the supernovae are no
longer orthogonal functions).

\begin{figure}
\begin{center}
\includegraphics[width=0.65 \textwidth]{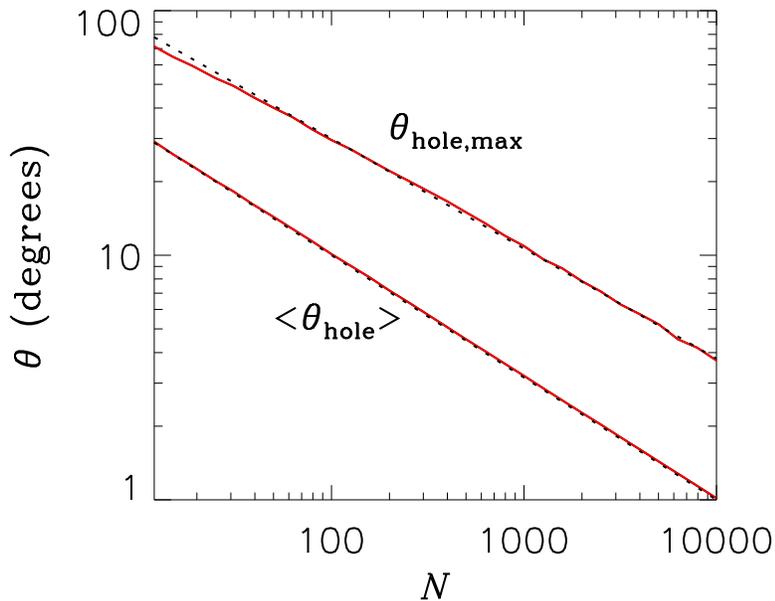}
\caption{The area-averaged size of holes ($\langle \theta_{\rm hole}
\rangle$), and the size of the largest hole ($\theta_{\rm
hole,max}$) on the sky as a function of the number of supernovae are
shown as full (red) lines. Overplotted are dotted lines
corresponding to $\langle \theta_{\rm hole} \rangle \propto
N^{-1/2}$ and $\theta_{\rm hole,max} \propto
N^{-0.45}$.}\label{fig:holes}
\end{center}
\end{figure}

As noted in \cite{Haugboelle:2006uc} a very important quantity is the size
of the largest holes in the distribution of supernovae on the sky.
We define a hole at a given point $p$ on the sphere as the largest
circle, which can be drawn around that point without encompassing any
supernovae. The hole can be quantified in terms of the angle
$\theta(p)$ of that sphere. For a random point distribution on the
sky the area-averaged size of holes must then be proportional to
$N^{-1/2}$ where $N$ is the number of supernovae in a given redshift bin. In
\fig{fig:holes} we show that this relation holds exactly when
averaged over many realisations. However, the size of the largest
hole drops as roughly $N^{-0.45}$ (as can be seen from
\fig{fig:holes}). In the region of interest this quantity is
also a factor 3-4 larger than the average hole size.

Since the largest holes are most problematic in the sense of loss of
orthogonality of the spherical harmonics it would be highly
desirable with a survey strategy, which attempts to minimise the
largest holes rather than the average size of holes. Quantitatively
this means a distribution on the sphere, which is as homogeneous as
possible for a given number of supernovae.

The survey telescopes supply us with light curves for a set of supernovae,
for a subset of host galaxies known redshifts can be found in galaxy
catalogues and the literature, and then the question is which of the
remaining host galaxies to measure redshifts for, to best reconstruct
the radial velocity field. Because we need the redshift of the host galaxy
and not the supernova itself, the observing strategy can be decided at a later
time when a large sample of supernova have already been observed
and the angular distribution is known.

We have tested different geometrical point measurements of radial
velocity fields with a known angular power spectrum, and found that
a regular homogeneous point distribution, such as a Healpix
distribution \cite{Gorski:2005}, consistently reconstructs
the power spectra with smallest errors. Unfortunately
nature is not regular, but given the realities of a semi-random
distribution of $N$ supernovae on the sky in each redshift bin to draw
from, it is possible to make an almost regular distribution,
containing only $N_{stop}$ supernovae, where typically $N_{stop} \sim
N/4$.

\begin{figure}
\begin{center}
\includegraphics[angle=90,width=0.48 \textwidth]{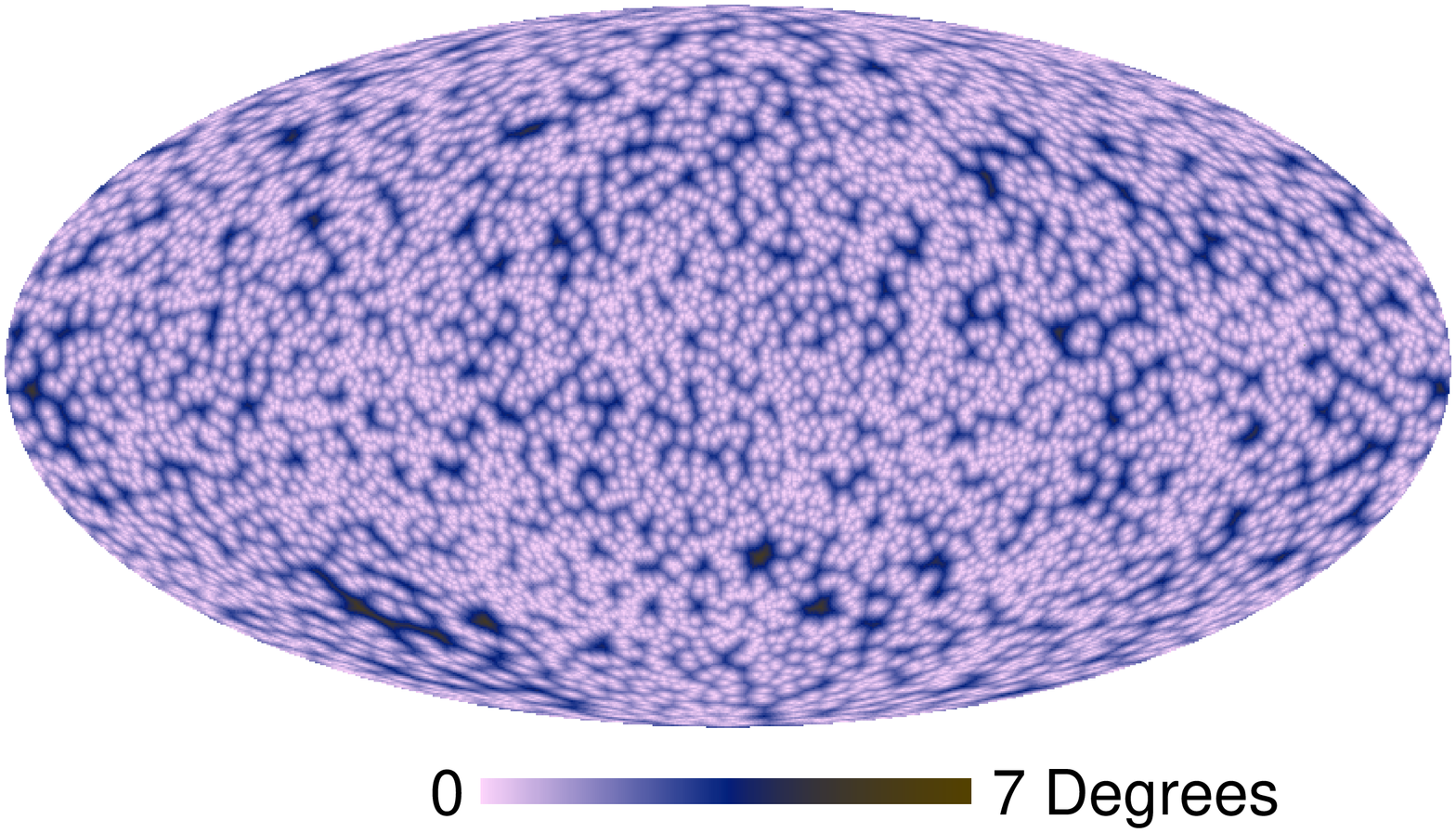}
\includegraphics[angle=90,width=0.48 \textwidth]{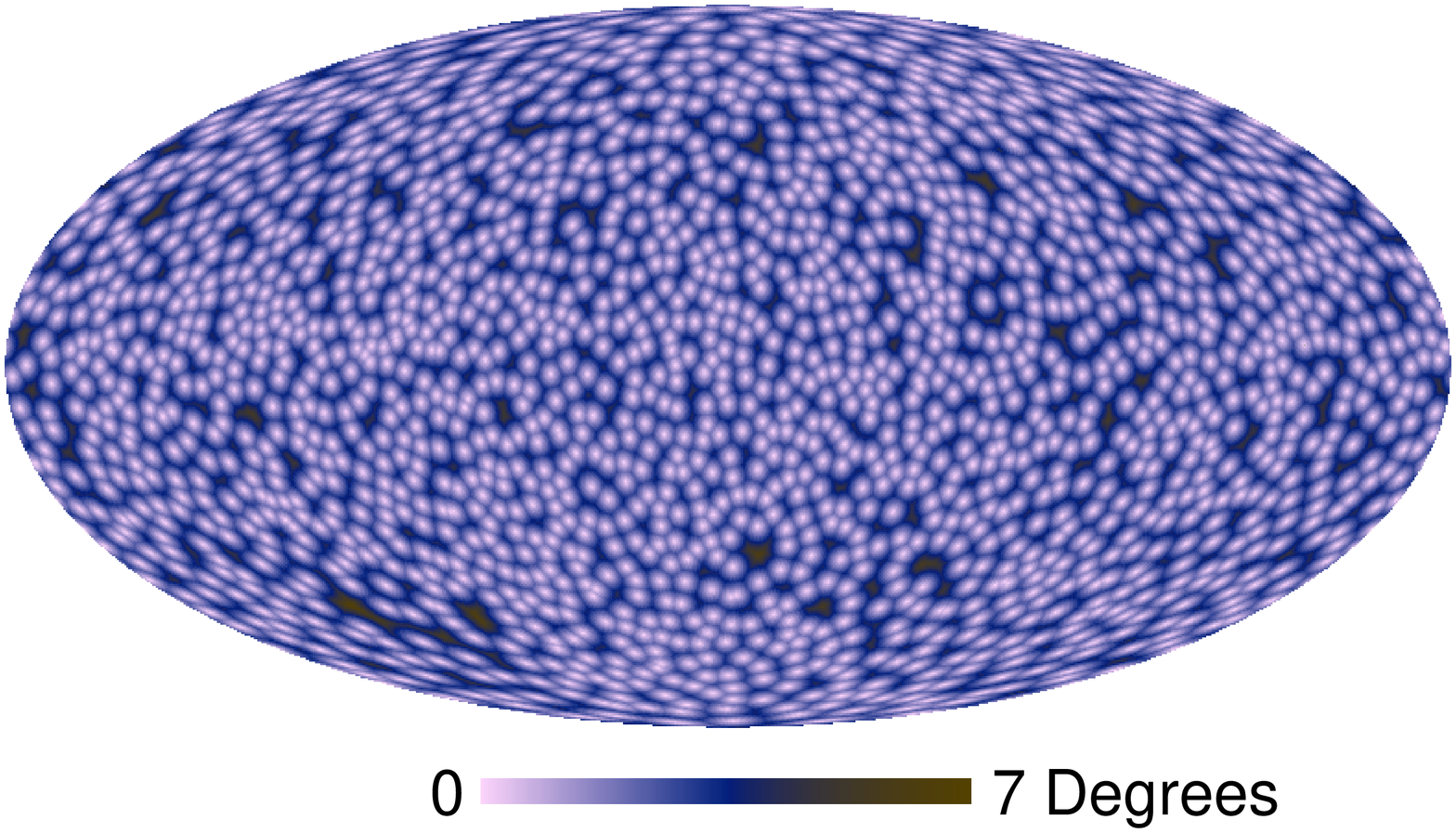}
\caption{The angular distribution of holes for 7125 random supernovae (left) and for 1781 glass supernovae drawn
from the 7125 supernovae using a basis of 445 random supernovae (right). The regular distribution of the glass
supernovae is apparent, while the random supernovae follow the underlying density field.}\label{fig:glassdist}
\end{center}
\end{figure}

\begin{figure}
\begin{center}
\includegraphics[width=0.8 \textwidth]{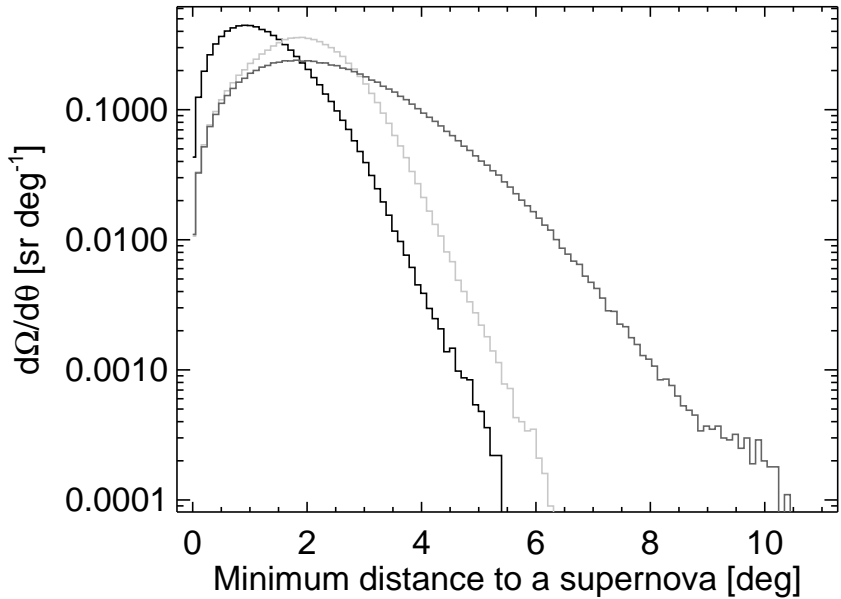}
\caption{Histogram of the angular distribution of holes for 7125 random supernovae (black curve),
1781 random supernovae (dark grey), and 1781 glass supernovae drawn from the 7125 supernovae (light grey).
The number of large holes in the case of a glass distribution is almost the same as for
the parent random distribution.}\label{fig:glasshist}
\end{center}
\end{figure}

To test our algorithm, we have first made mock catalogues
of supernovae from an N-body simulation, by selecting an observer at random,
but following the mass distribution; and then selecting $N_{tot}$ supernovae
at random. The probability of selecting a given supernova is proportional
to the matter density in the bin. Then our algorithm proceeds
as follows -- applied to the separate redshift bins of the individual mock catalogues:
\begin{itemize}
\item{A random subset containing $N_{start}$ of the points are drawn from the
parent distribution of $N$ supernovae in the redshift bin. They correspond to host galaxies with known redshifts.}
\item{The supernova, outside the subset, with the maximal minimum angular distance to the
already selected supernovae is added to the subset (a redshift is measured for the host galaxy of this supernova).}
\item{The last point is iterated adding new supernovae until $N_{stop}$ of the points have been selected.}
\end{itemize}
The resulting distribution is homogeneous but irregular, resembling
a glass distribution (see \fig{fig:glassdist}). The distribution of
holes, or, more precisely, the minimum distance from a given point
on the sky, to a host galaxy, for a set of glass supernovae, compared to a
random set, is characterised by having approximately the same
average distance to a supernova, but no big holes in the angular
distribution. Using only a quarter of the supernovae, it is competitive
with the parent random distribution. (see \fig{fig:glasshist}, and the
variance on the geometric error in \fig{fig:fig3}).

\subsection{Power spectra derived from a combination of mock observations and catalogues}
To estimate how precisely cosmological parameters can be
measured we will use three different datasets in the following:
\begin{itemize}
\item{A ``full'' dataset containing $N_{tot}=92,100$ supernovae, distributed
randomly according to mass in the N-body simulations. This
corresponds to three years of full sky coverage by survey
telescopes, with follow up observations on the redshift of all host
galaxies. The maximum redshift is $z=0.13$, corresponding to $400\, \hMpc$
and the data is analysed in six redshift bins.}
\item{A ``glass'' dataset of $N_{tot}/4$ supernovae, constructed from the
full dataset using 25\% of the supernovae, and the algorithm described above in this section.}
\item{A ``small'' dataset which contains $N_{tot}/4$ supernovae, distributed
according to mass in the N-body simulations.}
\end{itemize}

\begin{figure}
\begin{center}
\includegraphics[width=0.8 \textwidth]{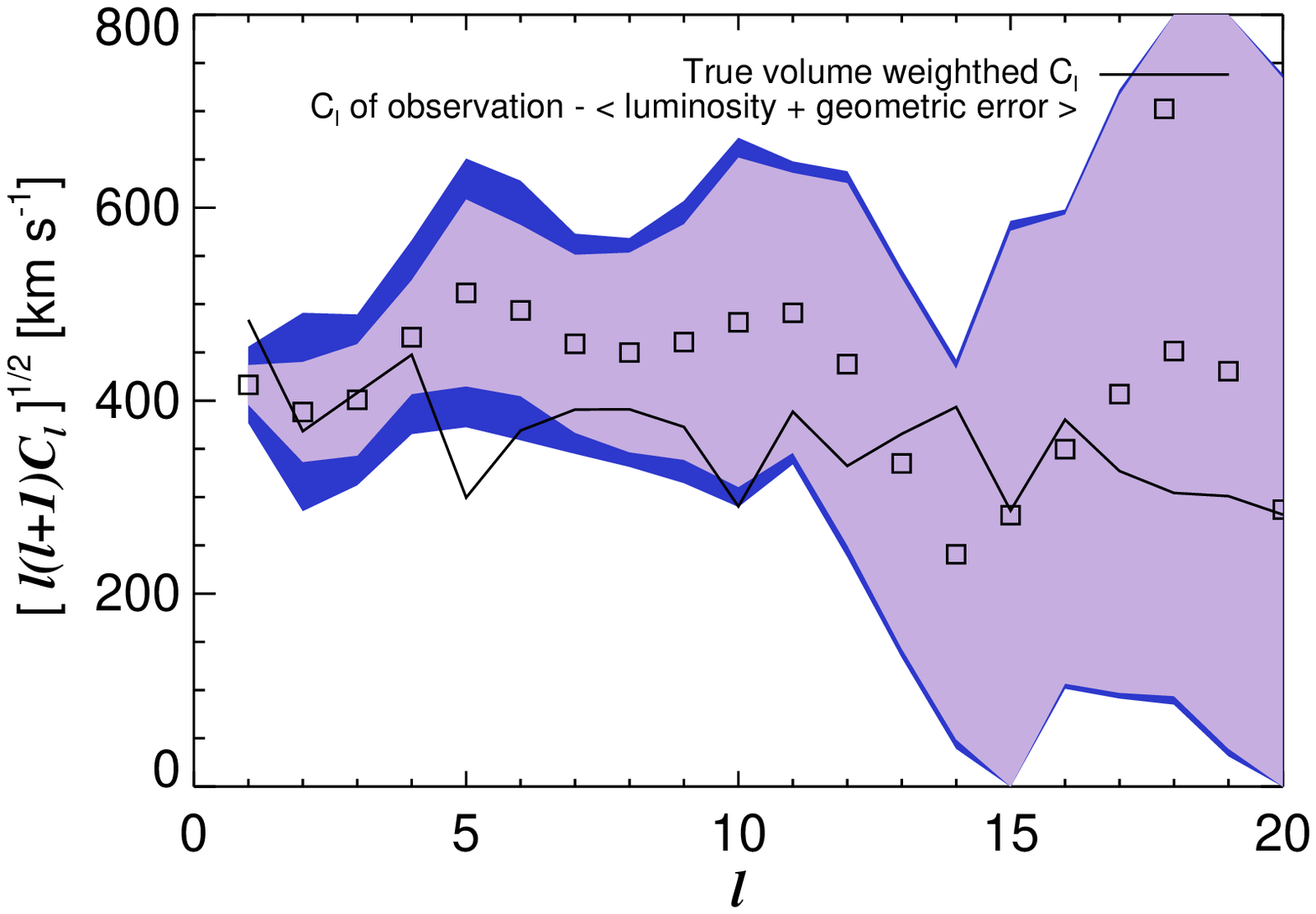}\\
\includegraphics[width=0.8 \textwidth]{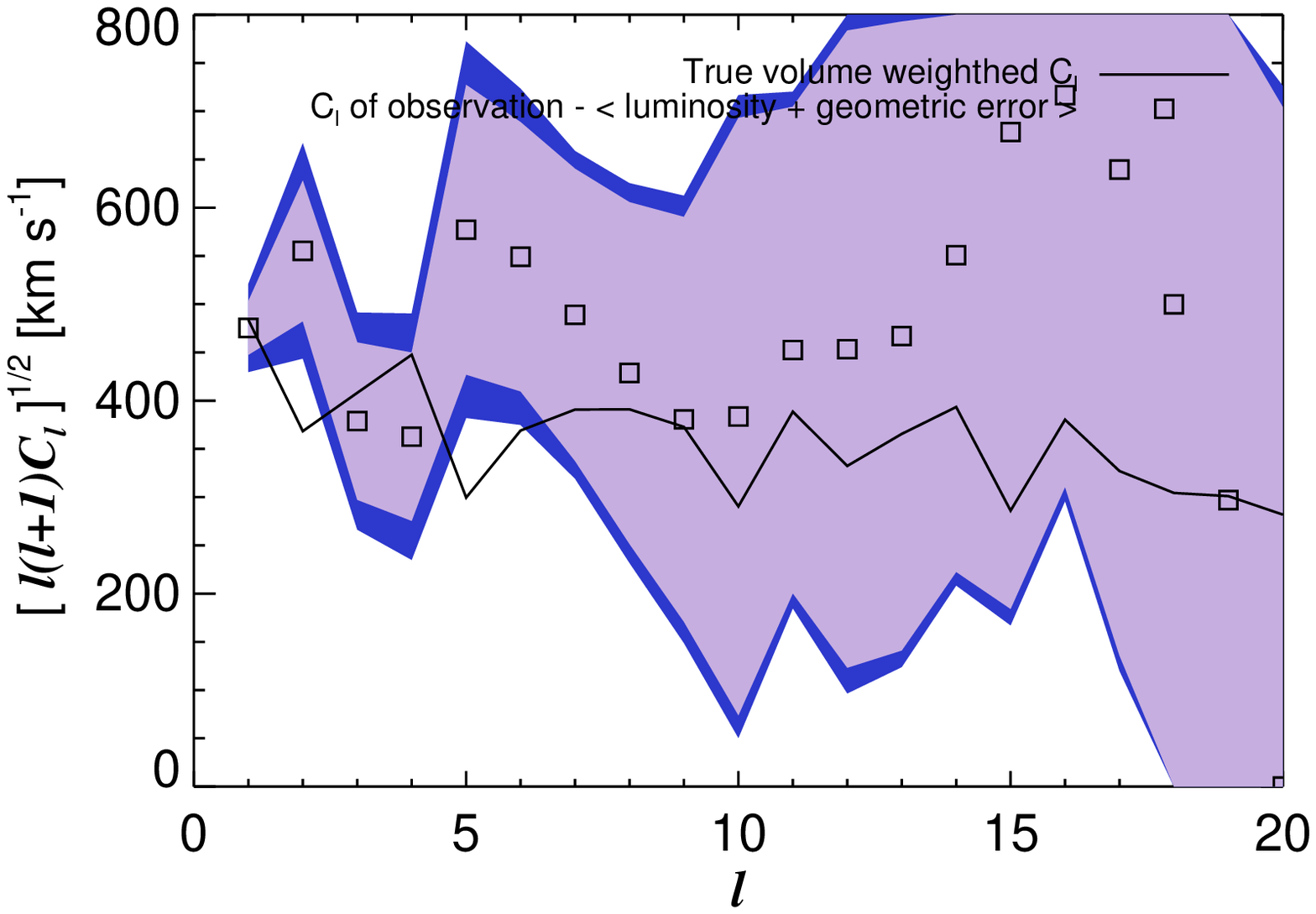}
\caption{The radial velocity angular amplitude spectrum in a bin centered at a comoving distance
of $150 \,\hMpc$ based on a synthetic survey bin with 1781 glass (random)
supernovae is shown at the top (bottom). Blue (purple) indicate the errors due to
the noise correction procedure and including (excluding) cosmic
variance with an estimated luminosity scatter of $\Delta m = 0.08$ on each
supernova. The full line is the true volume averaged amplitude spectrum for the
specific observer, while the squares show the corresponding
synthetic observation.}\label{fig:1781SNe}
\end{center}
\end{figure}

Using the average power spectra of both the geometric and the luminosity
errors to correct the power spectrum of a single mock observation,
we are able to probe the power spectrum up to $l_{max} \sim 20$ at
distances of $z=0.03-0.13$ (see \fig{fig:RNG7125_149}) quite robustly,
when the full data set is included in the analysis.
In \fig{fig:1781SNe} is shown the equivalent angular power spectra
using the glass and the small data set. It is clear that both the
overall quality and the systematic of the angular power spectrum are
greatly enhanced by using a glass distribution. Notice in \fig{fig:fig3},
how the variance in the geometric error of the glass supernovae is comparable to that
of the parent full data set, and at the same level as cosmic variance, while
the variances in both the geometric and the luminosity error from the small set is
significantly higher for $l>8$ and $l>12$ in the redshift bin centered at
$100\,\hMpc$ and $150\,\hMpc$ respectively.

It should be noted that the reason why this "glass" distribution
works so well is exactly because it is close to being homogeneous,
i.e.\ it has no large holes.

\section{Cosmological parameters}\label{sec:cosmo}
From our mock survey out to $z=0.13$ we have tested the ability to
probe the parameters $\Omega_m$ and $\sigma_8$ which are directly
related to the shape and amplitude of the matter power spectrum
around the peak. Specifically we obtained our best model estimate
of the observed radial velocity angular power spectrum
$C_l^m(\sigma_8,\Omega_m)$, including cosmic variance, geometrical
sampling noise, and the luminosity scatter of the supernovae,
as the average of 900 supernova mock surveys performed by 40 differently
located observers. As the probability density of $|a_{lm}|^2$
is exponential ($\chi^2$ with 2 degrees of freedom) the $\log$
likelihood function for a given mock survey within a specific
redshift bin is given by
\begin{equation}
 -\ln L(\sigma_8,\Omega_m) =
  \sum_{l=1}^{l_{max}}(2l+1)\left[\ln C_l^m(\sigma_8,\Omega_m) +
  C_l/C_l^m(\sigma_8,\Omega_m)\right]
\end{equation}
where $C_l$ is the calculated angular power spectrum for
the specific redshift shell. The total likelihood function
is obtained as the product of the likelihood functions for all
redshift bins.

\begin{figure}
\begin{center}
\includegraphics[width=0.65 \textwidth]{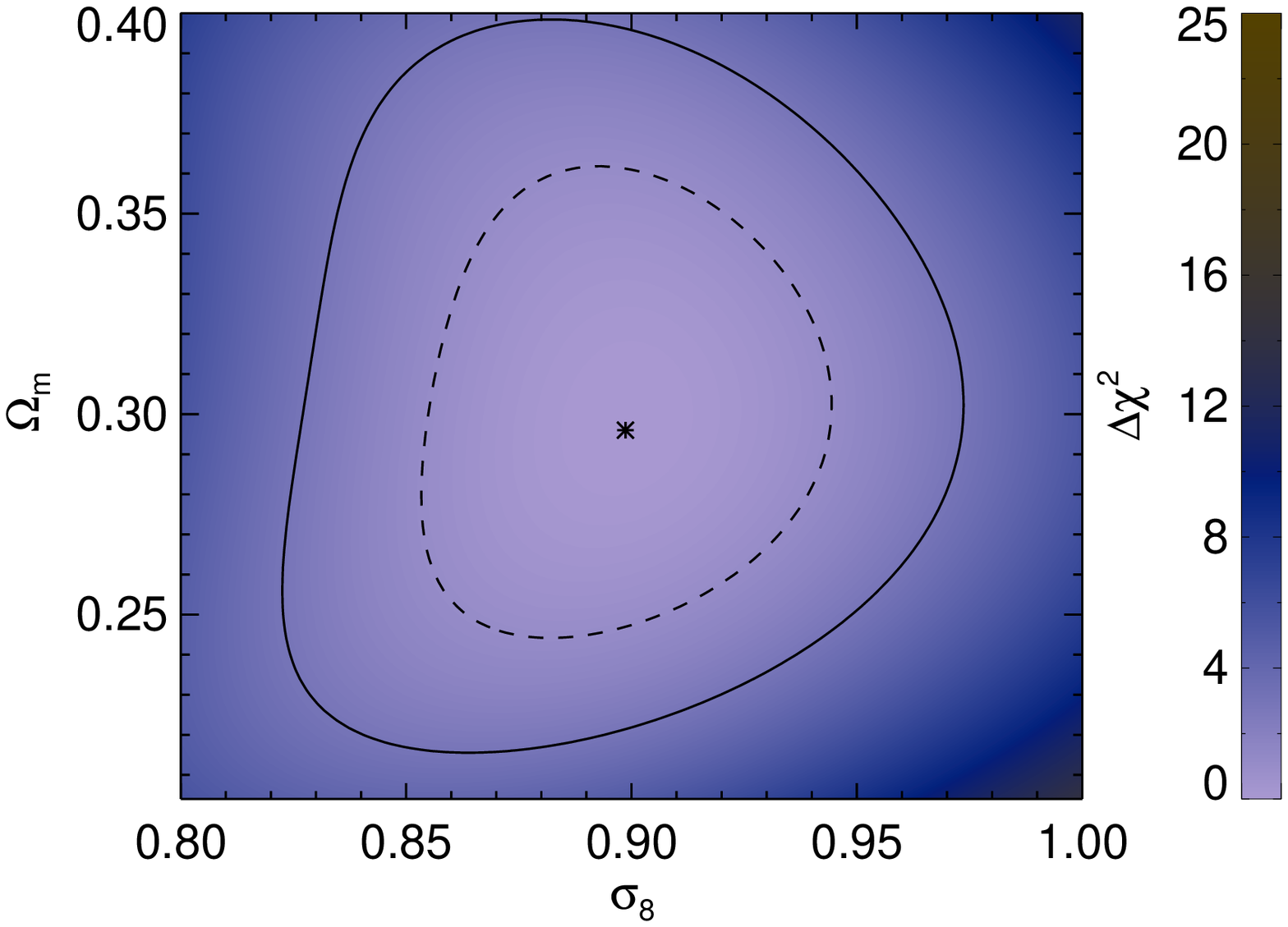}\\
\includegraphics[width=0.65 \textwidth]{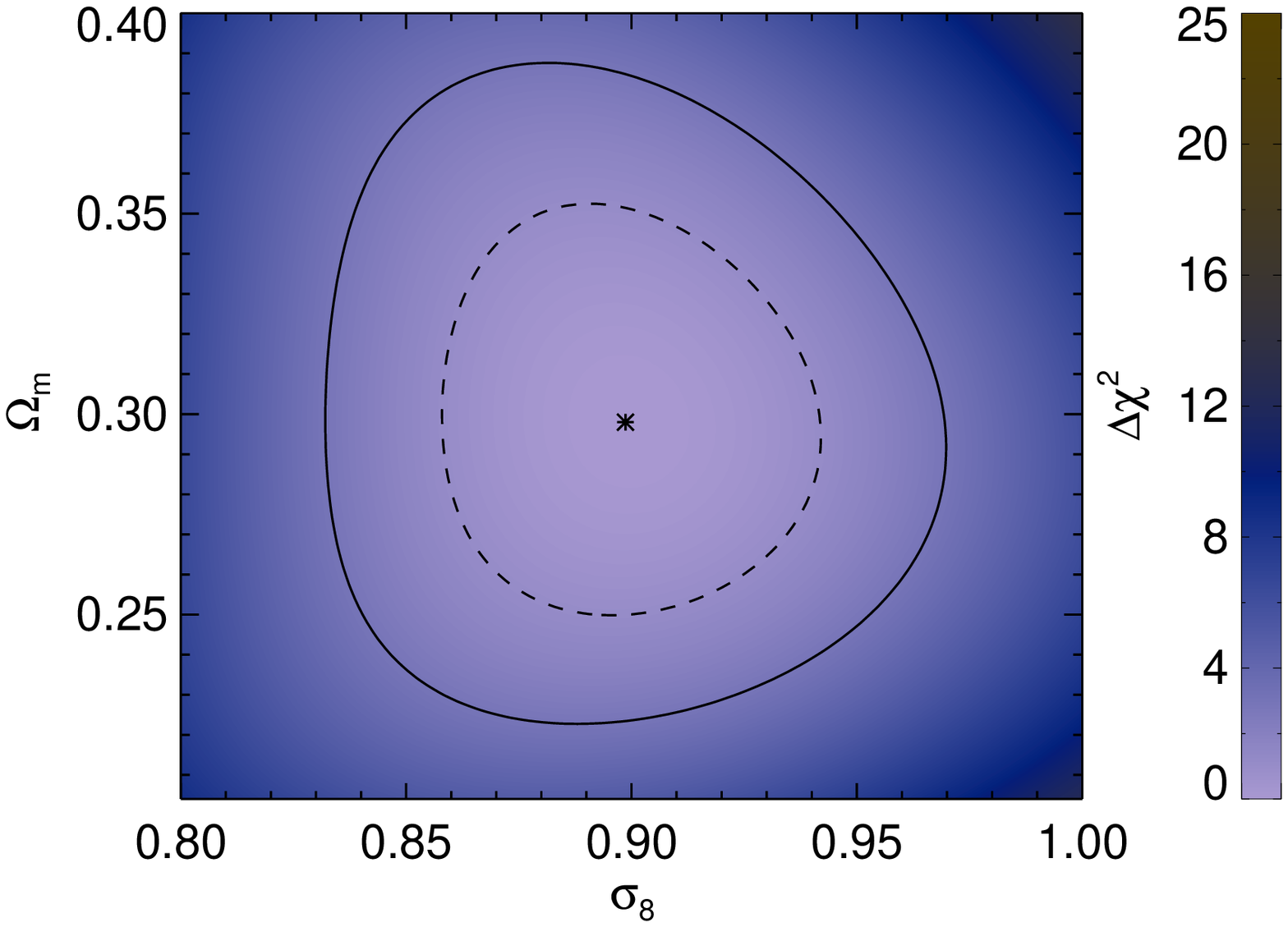}\\
\includegraphics[width=0.65 \textwidth]{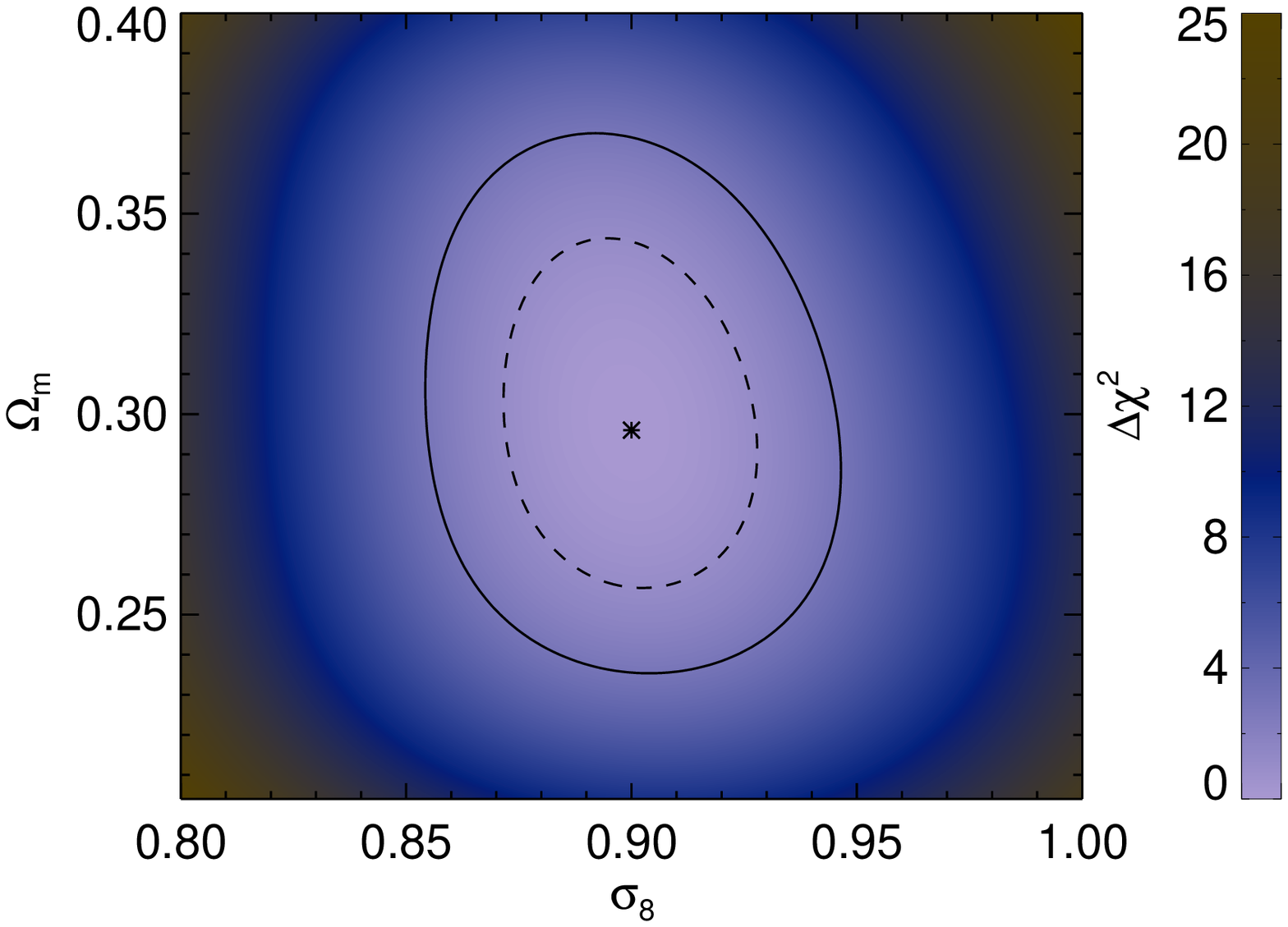}
\caption{The 2D 68\% and 95\% C.L. contours for $\sigma_8$ and
$\Omega_m$ for the three synthetic datasets: small (top), glass
(middle), and full (bottom).}\label{fig:2Dcontour}
\end{center}
\end{figure}

\begin{figure}
\begin{center}
\includegraphics[width=0.48 \textwidth]{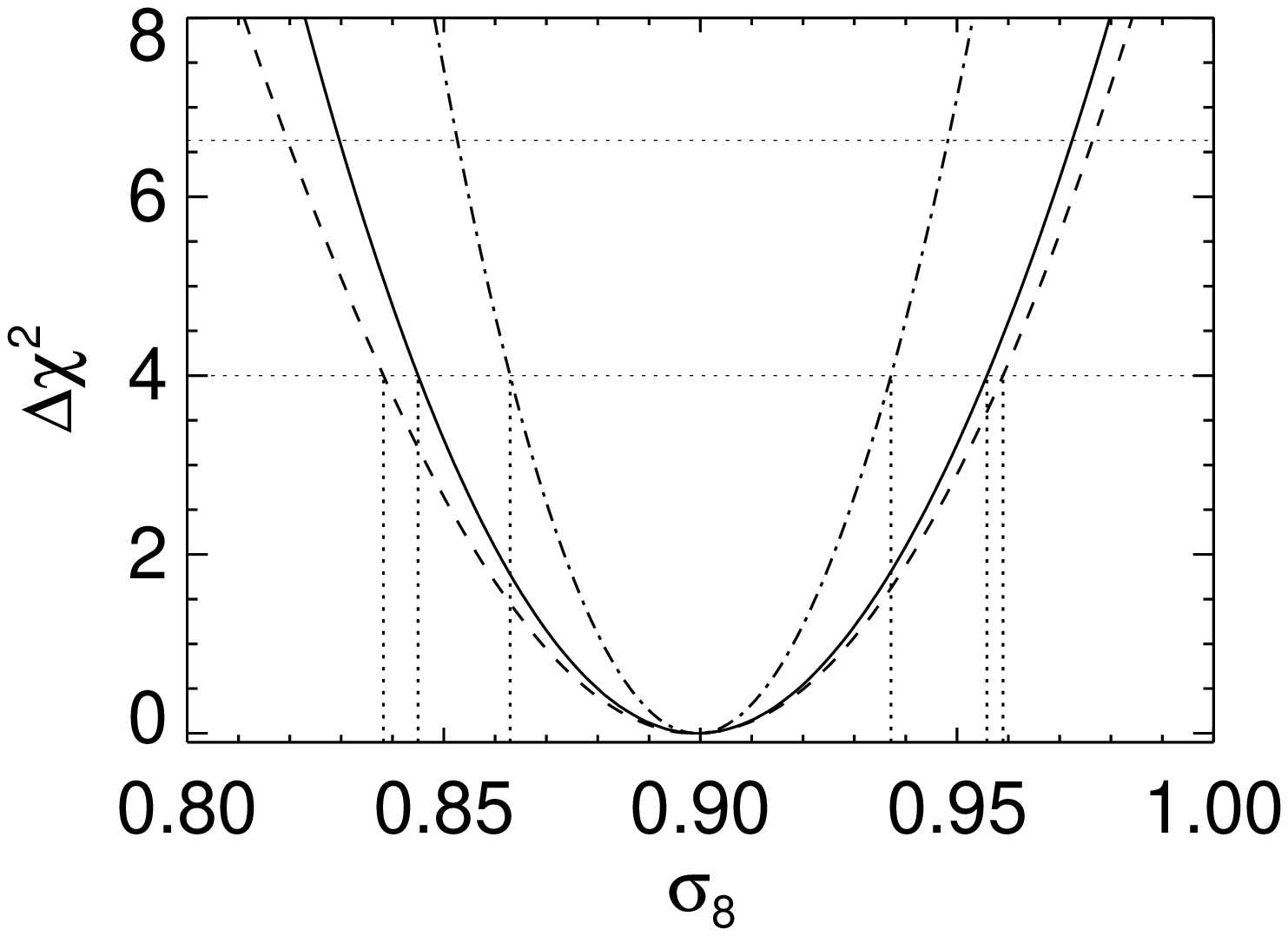}
\includegraphics[width=0.48 \textwidth]{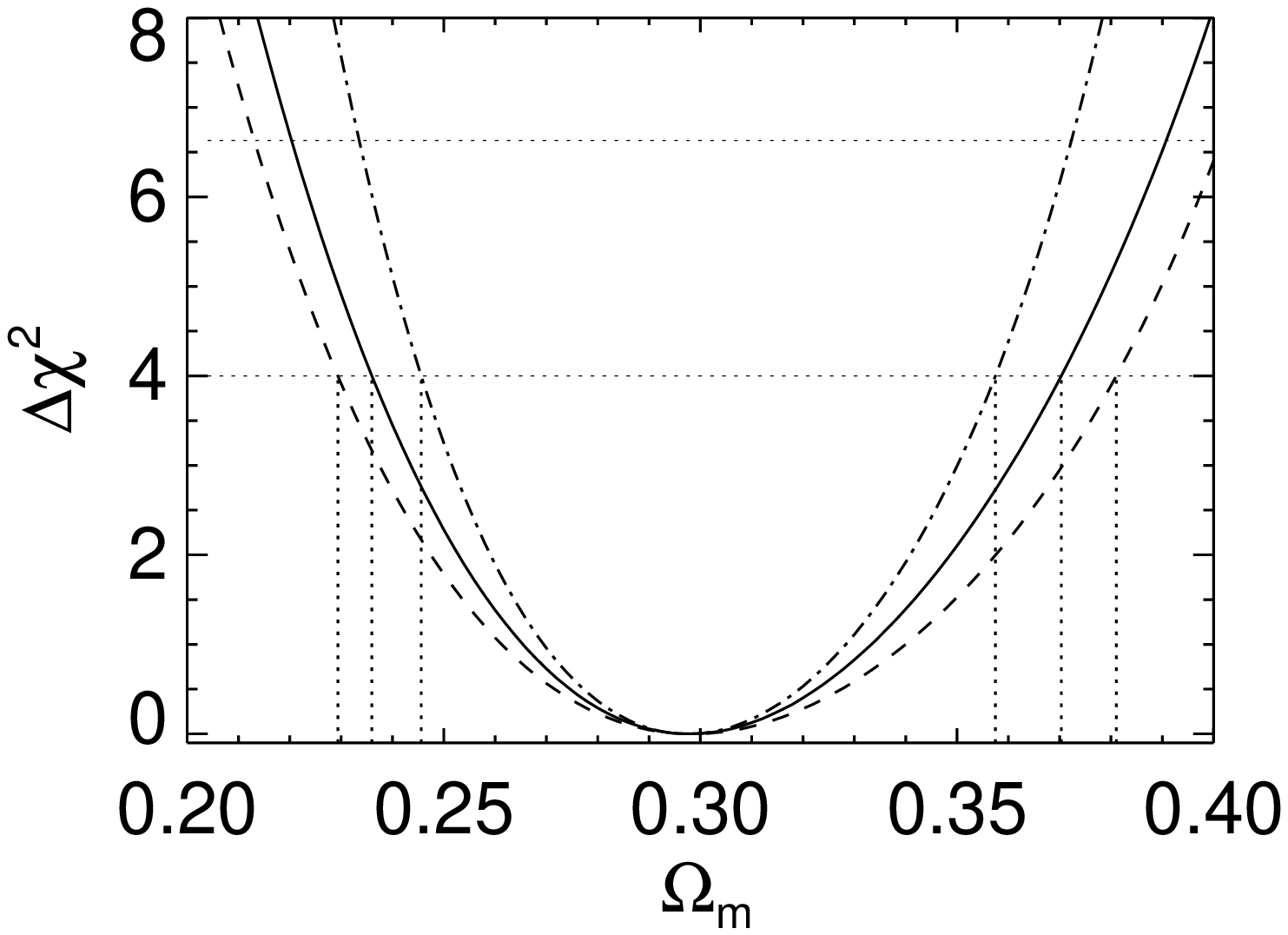}
\caption{The 1D $\Delta \chi^2$ for $\Omega_m$ (left) and $\sigma_8$
(right) for the three different datasets.}\label{fig:1Dcontour}
\end{center}
\end{figure}

In \fig{fig:2Dcontour} we show 2D contours of the likelihood
function based on a fiducial model with $\Omega_m=0.3$,
$\sigma_8=0.9$. In \fig{fig:1Dcontour} we show the
corresponding 1D likelihood function marginalised over the other
parameter.

With the small dataset a precision on $\sigma_8$ of roughly 0.06 can
be achieved, and with the glass data this is somewhat improved. It
should be noted that although the difference between the ``small'' and
the ``glass'' data sets is quite small in the present case it would be
much larger if the intrinsic scatter, $\Delta m$, could be reduced, or
there were fewer supernovae in the sample.
As supernovae become better calibrated standard candles the
importance of the ``glass'' distribution will increase accordingly.

With the full data set the precision is roughly 0.03 at 95\% C.L.
This is comparable to the estimated precision for Planck+LSST weak
lensing \cite{Hannestad:2006as}. However, we stress that a more
thorough parameter study should be performed for the supernova data
before a direct comparison can be made with other future probes.

The cosmological parameter estimates can be improved significantly
by either reducing the intrinsic scatter, $\Delta m$, or increasing
the survey volume. Going to $z = 0.2$ is doable in the sense that
the changes in apparent magnitude of supernovae are still dominated by
peculiar velocities compared to lensing up to fairly large $l$
\cite{Bonvin:2005ps}. In this case the total survey volume will be
of order $V \sim 0.9 h^{-3}$ Gpc$^3$. This is comparable to the
SDSS-LRG effective volume of $V \sim 0.75 h^{-3}$ Gpc$^3$ \cite{sdsslrg} because
the supernova survey is full-sky, and the number of objects will be larger
than the number of measured LRG galaxies.

\section{Discussion}
We have performed detailed numerical studies of how future type Ia supernova
surveys can be used to probe the large scale velocity field of the local universe
out to redshifts $z \sim 0.13$. Our approach was to use high resolution N-body
simulations of structure formation. From these simulations we created many synthetic
data sets and used a realistic analysis pipeline to extract the angular
velocity power spectrum at different redshifts for each data set.

These velocity power spectra are directly related to the underlying
matter power spectrum because the velocity flows are caused by the underlying
gravitational potential. Indeed, one of the main advantages of the presently
proposed technique is that it directly probes the gravitational potential, unlike
galaxy surveys that measure the galaxy power spectrum. This quantity is related to the
matter power spectrum only via a scale dependent bias parameter.
In this regard future large scale measurements of the velocity power spectrum using
type Ia supernovae are similar to the future large scale weak lensing measurements, which
also probe the gravitational potential directly. Both probes are complementary to
galaxy surveys and combining either with a large scale galaxy survey offers a powerful method for
extracting the bias parameter. It will indeed be very interesting to investigate
the interplay between future large scale galaxy surveys and supernova surveys in more detail.

Because the velocity power spectrum measured with type Ia supernovae is directly
related to the underlying matter power spectrum it is particularly useful for
probing the amplitude of fluctuations on scales of a few hundred Mpc, usually quantified by the
parameter $\sigma_8$. Using synthetic data sets constructed for different cosmological
models we performed a simple likelihood analysis based on a fiducial model with $\Omega_m=0.3$,
$\sigma_8=0.9$, and found that $\sigma_8$ could realistically be measured with a precision
of 3-5\% (95\% C.L.). This is comparable to the estimated precision of future weak lensing
measurements. Furthermore, the precision can be improved significantly by either
increasing the survey volume or reducing the intrinsic uncertainty of each supernova.

Much work is still needed in order to accurately gauge the
potential of future large scale surveys of low-redshift supernovae
for cosmology. However, the present study clearly shows that such
surveys, with a relatively small observational effort, can be competitive
with other cosmological probes, which will be available in the next decade.

\ack We acknowledge many useful discussions with Tamara Davis,
Ariel Goobar, and Edvard M{\"o}rtsell. We thank the Danish Centre
of Scientific Computing (DCSC) for granting the computer
resources used. T.H.~thanks the Dark Cosmology Centre for hospitality
during the course of this work. Some of the results in this paper have been derived
using the HEALPix \cite{Gorski:2005} package.

%%%%%%%%%%%%%%%%%%%%%%%%%%%%%%%%%%%%%%%%%%%%%%%%%%%%%%%%%%%%%%%%%%%%%%
\section*{References} %%%%%%%%%%%%%%%%%%%%%%%%%%%%%%%%%%%%%%%%%%%%%%%%
%%%%%%%%%%%%%%%%%%%%%%%%%%%%%%%%%%%%%%%%%%%%%%%%%%%%%%%%%%%%%%%%%%%%%%

\end{document}